# Development of a culturally-oriented website usability evaluation
# (Research in progress)

| | |
|---:|:---|
| Journal: | *15th Americas Conference on Information Systems* |
| Manuscript ID: | AMCIS-0676-2009.R1 |
| Submission Type: | Paper |
| Mini-Track: | Interface Design, Evaluation, and Impact < HCI Studies in Information Systems (SIGHCI), Cultural Issues and Information Technology Diffusion < Diffusion of IT (SIGADIT) |
| | |

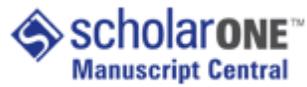

# Development of a culturally-oriented website usability evaluation

**Susana B. Vidrio-Baron**
*Iowa State University*
*sbvidrio@iastate.edu*

**Andrew W. Luse**
*Iowa State University*
*andyluse@iastate.edu*

**Anthony M. Townsend**
*Iowa State University*
*amt@iastate.edu*

**ABSTRACT**

As the uni-cultural studies of website usability have matured, the paucity of cross-cultural studies of usability become increasingly apparent. Moving toward these cross-cultural studies will require the development of a new tool to assess website usability in the context of cultural dimensions. This paper introduces the preliminary results from the first phase of this project and then presents the proposed method for the research in progress that specifically is directed to the development and quantitative evaluation of a measurement scale of a culture sensitive measurement of website usability. The recognition of the need to develop this scale resulted from the identification of culture-related shortcomings of previous measurement tools that have been used widely within the Management of Information Systems (MIS) literature.

**INTRODUCTION**

It has been widely acknowledged that there is a need to conduct more culturally-oriented website usability evaluation. Sears (2000) conducted a thorough literature review within Human Computer Interaction (HCI) journals (i.e. International Journal of Human-Computer Interaction, Human-Computer Interaction) and was able to identify some previous efforts that aimed to provide practical insights and recommendations for additional research on the relation between cultural differences and design for the WWW.

Amongst the more recent definitions regarding culture and information technology, Aykin (2005) proposes a taxonomy of approaches to the understanding of globalization and internationalization. He proposes some of the following definitions obtained from the Localization Industry Standards Association (LISA) in the Localization Industry Primer (LISA, 2001):

> *"Globalization: The general process of worldwide economic, political, technological, and social integration.*
>
> *Internationalization: The process of ensuring, at a technical/design level, that a product can be easily localized. Internationalization is thus, part of globalization.*
>
> *Localization: The process of modifying products or services to accommodate differences in distinct markets. Localization makes products or services usable in, and therefore acceptable by, target cultures.*
>
> *Translation: The linguistic component of localization. It is choosing the appropriate text in a target language to convey the proper meaning from a source language text.*
>
> *Locale: That part of the user's environment that is dependent on language, country/region and cultural conventions."* (Aykin, 2005, pp. 4-5)

The ongoing discussion within the academic community regarding either globalizing or localizing the components of a website is at this point is mostly speculative, as research has yet to be done on the ways that localized and globalized web services work across cultural boundaries. As stated in the definitions provided previously, globalizing involves the development of standardized, supranational and, in a way, anonymous interfaces or systems (in the sense that no particular nationality, credo or particular local trait can be identified). Within the HCI terminology this tendency has been identified as universal, user-centered design. As one might imply from the term, the main focus of this form of design is directed towards the identification of universal user characteristics that need to be addressed when designing human interfaces.

Conversely, other discussion has been oriented towards the recognition, identification and inclusion of cultural values, dimensions, elements or traits that are specific and particular to determined countries, regions or even languages, along with some prognosis on how these characteristics affect appropriate design. A number of disciplines (i.e. Management, Marketing, and Design among others), have developed theoretical frameworks derived from a more classical treatment of the cultural construct. This is the case of the work conducted by the Dutch sociologist Geert Hofstede and son (Hofstede & Hofstede, 2005) in the latest verison of the cultural dimensions and values research conducted by Geert Hofstede with the support of IBM in 53 countries around the world. Originally, the Values Survey Module (VSM) instrument that he developed was created in order to identify probable cultural dimensions within the organizational environment. Eventually, these dimensions were additionally used to address the identification of cultural dimensions within the national environment as well as for the individual/personality unit.

These dimensions (i.e. Power Distance, Uncertainty Avoidance, Indivudalism-Collectivism, Masculinity-Femininity, and Long-Term/Short-Term Orientation) have been widely utilized across multiple disciplines, including the IS discipline. Leidner and Kayworth (2006) found that the dimensions of culture developed by Hofstede were frequently used in research, consistently treated as independent variables. Clearly, within the IS literature, the broad cultural construct has been thoroughly empirically tested, and thus would appear ripe for application to cross-cultural usability models.

**Literature Review**

In a broad spectrum of research, system design and user satisfaction variables have been a frequent component of usability models. We would note that that the way in which HCI usability assessment has been conducted differs from the treatment given by the IS discipline. That is, for the HCI and Design literature, there has been a recognized call to conduct empirical validation of some of the "cultural difference" propositions in order to address the need to embed the cultural dimensions into the method that usability evaluations are being conducted, particularly within the assessment of website usability. Website usability, because of its dependence on embedded content, is much more culturally aligned than other forms of application usability that do not present both content and technical structure.

When we say that there is a need to conduct empirical validation of some propositions, this reflects the work work that has been conducted from workshops conducted around mainly the United States by Aaron Marcus. Marcus is one of the founders of the Usabilty Professionals Association (http://www.upassoc.org/) and has been working towards the promotion and adoption of usability services througout the world.

Marcus and Gould (2000) have adopted the cultural dimensions proposed by Hofstede in order to identify the implications that each dimension has in explaining website design or to identify them by assessing a website. Mainly, the result of this effort has the form of propositions, considerations, recommendations or research questions.

Amid some of the website elements consistently reported by Marcus and Gould, the following can be identified: (1) acces to information, (2) hierarchies in mental models, (3) information recognition, (4) security, (5) graphics, sounds, animations, colors and images, (6) communication style and (7) navigation.

Another important contribution to this analysis is the work conducted by Barder and Badre (1998); they coined the term *"culturability"* which implies the artificial symbiosis of the usabilty evaluation of websites and the particular cultural elements that can be identified within them. They also identify those particular cultural elements as *"cultural markers"* and provide the following definitions for clarification:

> *"Cultural Marker: Cultural markers are interface design elements and features that are prevalent, and possibly preferred, within a particular cultural group. Such markers signify a cultural affiliation. A cultural marker, such as a national symbol, color, or spatial organization, for example, denotes a conventionalized use of the feature in the web-site, not an anomalous feature that occurs infrequently.*

*Vidrio-Baron et al.*                    *Development of a culturally-oriented website usability evaluation*

> *Culturally Deep vs. Shallow Sites: We define a culturally deep web-site as one that occurs in the native language of its country of origin and links to other native-language sites. A culturally shallow site is one that occurs in a secondary language and links to other secondary language sites."[1]*

Finally, the more recent work conducted by Singh and Pereria (2005) will be presented. It resulted in the only book entirely devoted to the development of an empirical rationale for cultural customization within electronic commerce websites. Of particualr interest for the proposed research effort are the resulting operationalized cultural dimensions (the ones proposed by Hofstede) within the context of electronic commerce webpages. After the literature review, it became apparent that this was the first attempt conducted in order to provide a more empirical starting point that could help future research initiatives when assessing websites in lieu of the cultural dimensions context.

For their analysis, the authors included four of the cultural dimensions proposed by Hosftede, namely Power Distance, Uncertainty Avoidance, Individualism-Collectivism, and Masculinity-Femininity. Additionally, the authors integrated one of the variables previously proposed Hall (1976) which is High-Low Context. In addition to Marcus and Gould, Singh and Pereira developed a set of propositions, but are in no way intending to test them with more quantiative methods.

Basically, the actual literature development has reached a peek at the development of propositions and the instrumentation of possible items that aim to measure culture on websites. These propositions have been the result mainly of a qualitative and iterative process conducted in recent years. One interesting finding from the review of this literature is the consistency of the propositions obtained and reviewed from these different sources.

Since the literature review resulted in a diverse and eclectic compilation of the main usability evaluation methods, techniques, and tools; and, since all of them are derived from very different (yet interconnected and complementary) disciplines, we aim to present a preliminary and exploratory methodological proposal to include the cultural dimensions when assessing websites' usability.

In the following section, we will describe the process we have followed based on the previously conducted analyses. This process, which has not been completed, has given some insights and provided preliminary information that will lead to the proposed following effort that will aim to provide a method to include culture within the website evaluation process.

**PROPOSED METHOD AND RESEARCH IN PROGRESS**

The first stage of the analysis included a website deconstruction that helped to understand the common sections or components that could be identified. Through the literature review, it became evident that when defining culture and technology, the issue of identifying them as related to the individual or the society must be addressed. Also, within the literature websites have been mainly identified as organizational or from corporations. So, for the scope of this analysis, electronic government websites will be treated as organizational or corporate websites. Regardless of whether we are exploring cultural effects on websites on commercial or governmental organizations, the key (as we see it) is to examine "like to like" types of sites that have an intense similarity of common purpose. Thus, bus companies, drivers' license bureaus, chambers of commerce, etc. have the same fundamental purpose regardless of its home country; these types of cites then make it easier to discern the cultural artifacts extant that create differences in site structure.

**Previously conducted analyses**

In order to understand how website design is performed and which elements designers consider necessary when constructing a website, a focus group was conducted with graphic arts students (most of them website designers or HCI students). The objective was primarily to validate if previously used metrics for website evaluation seemed to be exhaustive and culturally inclusive, and to identify if the focus group thought there would be a correlation among these elements and the four Hofstedes' dimensions of culture. Previously, we mentioned some of the most salient website usability evaluation elements found in the literature. After conducting some data collection from the designers, we could identify the following website elements: 1) Navigation, 2) Layout, 3) Design, and 4) Content.

---

[1] Extracted exactly from: http://zing.ncsl.nist.gov/hfweb/att4/proceedings/barber/index.html

For this particular research project and since the unit of analysis was the websites, convenience sampling or selection of participants was conducted by targeting Iowa State University Graduate Students. The respondents had to have some interest in graphic design or Human-Computer Interaction and be fluent in English and Spanish.

To conduct the proposed assessment, a survey has been developed after the concretion of a correlation matrix that was tested among participants from the first focus group, and the definition of usability criterion to be compared or contrasted against Hofstede's national cultural dimensions. Once the survey was structured, it will be completed by the subjects, who preferably are citizens from the selected countries (Mexico, Argentina, Chile, and Brazil). They would answer the questions by browsing through the selected web pages and identifying the items presented on the survey. Why these countries? Because they correspond to the 37, 39, 40, and 45 places of the 2008 Electronic Government Readiness Index developed by the United Nations. Also, these countries would be representative of the Latin culture cluster.

**Limitations and Considerations for the Study of Culture and Websites**

Due to the lack of response from the Latino community in the selected location, on the first stage of the analysis only 7 participants conducted the evaluation. In order to capture the attention of these participants, a broad call for participation was extended within a network focused primarily on providing information about recent and future events amid the Latino community.

So, for the second phase of the proposed analysis, some suggestions and modifications to the way the assessment is conducted will be conducted.

**Proposed new method**

For the second phase of this evaluation, recent suggestions have been proposed. First, within the MIS community and after a thorough discussion of the method previously used, a change on the participants and the websites evaluated was suggested. Additionally, since the resulting product from the first iteration was a survey, or at least a survey draft, the idea of testing the survey or what is usually identified as scale measurement was proposed.

From the literature review it became very clear that for this particular analysis, there is not a specific scale that can be used. The closest substitutes that were identified were combinations of two particular metrics that have been used previously within the realm of complementary disciplines, one of those was the WebQual. The WebQual is a metric derived mainly from the marketing research efforts and it has been tested to be a metric for web service quality. Quality, efficency and efficacy are some of the main variables that define the Usability construct. Eventually through some testing and adaptation, the WebQual scale becane the de facto standard within the industry to measure usability.

The other set of scales where the ones derived from the work conducted by Hofstede in order to assess the cultural construct. These two metrics were independent form each other and also were identified to be the most widely used to conduct usability website evaluations within the context of culture.

Hence, since there were several criticisms on the nature, scope and items included within these two metrics, the need to construct a new one that captured the latent variables proposed was identified. Therefore, for the second phase of this analysis, a scale construction and measurement is proposed.

**Scale Measurement**

The reason behind the scale construction and measurement is an attempt to provide a quantitative measurement of the abstract theoretical variable which in this case is cultural website usability. We will conduct for this matter both a reliability and a validity test, scale has validity if it properly represents the theoretical construct it is meant to measure. A scale has reliability if repeated measurements under the same circumstances tend to produce the same results.

The previous work conducted by Marcus, Gould, Singh and Pereira will be used as parts of the proposed items. Since the scope of the new proposed analysis is to develop and measure a survey, we will focus the attention towards a different set of participants.

**Participants**

For the second phase, the participants will be undergraduate students at Iowa State University MIS program. Because the scale, at the moment, is written in English, we will focus our attention towards native English speakers. In a proposed future third phase, the translation and subsequent validation of the scale is proposed as well but, for the moment, we will focus the attention towards the English version. A set of activities will be posted for the participants to get involved and get to know the website so that they can provide a posterior assessment by answering the questions of the survey.

**Websites**

Governmental websites are supposed to be a good interface that should aim to portray cultural values to the target population. Since the main target population is the American citizens, we are assuming that the communication, graphics, structure and design will be directed to the intended audience, which posses a set of very particular cultural values. Additionally, as suggested by MIS scholars, the inclusion of a set of diverse websites, intended for an American audience, but some developed by foreigner designers, will be included in the analysis to make it more powerful. Following with the original research attempt, we will include other federal governmental websites from countries that are in the first 10 places of the UNPAN e-readiness list, which include the following websites:

| Country | E-government websites | UNPAN e-readiness ranking |
|---|---|---|
| Sweden | http://www.sweden.gov.se/ | .9157 (1st place) |
| Denmark | http://www.denmark.dk/en | .9134 (2nd place) |
| Norway | http://www.regjeringen.no/en.html?id=4 | .8921 (3rd place) |
| United States | http://www.usa.gov/ | .8644 (4th place) |
| Netherlands | http://www.government.nl/ | .8631 (5th place) |
| Republic of Korea | http://www.korea.net/ | .8317 (6th place) |
| Canada | http://canada.gc.ca/home.html | .8172 (7th place) |
| Australia | http://www.australia.gov.au/ | .8108 (8th place) |
| France | http://www.premier-ministre.gouv.fr/en/ | .8038 (9th place) |
| United Kingdom | http://www.direct.gov.uk/en/index.htm | .7872 (10th place) |

**Table 1 Selected Websites**

**Instrumentation**

The proposed scale will be posted online; the actual scale contains 45 items, which would be incorporated into the digital version. These items resulted from a preliminary examination and focus group with website design experts. Most of the items were obtained from the work conducted by Marcus and Gould but also from the literature review of Hofstede's Cultural dimensions. Before submitting the final version, another round of revisions and probable additions of items will be conducted. Since it has been stated that for the development of scales some set of recommendations should be taken into consideration (i.e. the bipolar nature of the items, the phrasing and writing, the use of several items and then probable reduction) another set of revisions seem to be adequate.

Additionally, as proposed by MIS scholars, another set of constructs derived from Hall's cultural dimensions will be also included. The Low-High Context dimension that has been derived from the work conducted by Hall (1976) and that will be included in the proposed analysis is the more salient whithin the literature. This particular dimension is considered important because it refers especifically to the way infomration is used and interpreted within cultures. It has been difined as: *"High context cultures use more symbols and nonverbal cues to communicate, with meanings embedded in the sitautional context. Low context cultures are societies that are logical, linear, action-oriented, and the mass of the information is explicit and formalized"* (Singh & Pereira, 2005, pp. 55).

For the particular research project, a survey with a scale of titems being rated from 1 to 7, being 1 the Low level of agreement and 7 the high level of agreement will be developed by including some of the items that have been previously developed and some others added from the literature. Table 2 presents a list of the proposed items:

| Please select the number that best represents your impressions after browsing through the Website |
|---|
| 1. Is the information highly structured into many layers, hierarchies or specific sections? |

| | |
|---|---|
| 2. | Do you think there is redundancy of information (information being repeated)? |
| 3. | Do you think that there are many secured or unauthorized sections (login sections or secure sections)? |
| 4. | Do you think the web site lets you explore and gives you flexibility when navigating through it? |
| 5. | Do you think that graphics, sound and animation are used for utilitarian purposes? |
| 6. | Do you think the web site navigation scheme is intended to prevent citizens from becoming lost? |
| 7. | Do you think that there are many secured or unauthorized sections (login sections or secure sections)? |
| 8. | Do you think the web site lets you explore and gives you flexibility when navigating through it? |
| 9. | Do you think the web site navigation scheme is intended to prevent citizens from becoming lost? |
| 10. | Do you think there is a clear mental model and map that focus on reducing errors on the website navigation? |
| 11. | Do you think that graphics, sound and animation are used for utilitarian purposes? |
| 12. | Do you think the human figures tend to be more authoritarian? |
| 13. | Are the human images smiling? |
| 14. | Are their clothes formal? |
| 15. | Are the colors of the web site dark and solid? |
| 16. | Are children and babies present? |
| 17. | Are human images holding tools or machines? |
| 18. | Are the non human images related to artificial structures? |
| 19. | Are the non human images related to cartoons? |
| 20. | Are the non human images drawn in solid and dark colors? |
| 21. | Do you think the human figures pose is formal? |
| 22. | Do you think the attention to images is not focused on any person? |
| 23. | Do you think there are grand structures with full perspective and sky? |
| 24. | Do you think the buildings present on the website are big and tall? |
| 25. | Do you think that groups of people are present? |
| 26. | Do you identify any effort made by the government to address the elder and less favored citizens' needs? |
| 27. | Do you think that the web site provides free access to all the information (login sections or secure sections)? |
| 28. | Do you think there are many authorities and official symbols clearly identified? |
| 29. | Is the information structured into many layers, hierarchies or specific sections? |
| 30. | Do you think the government promotes patriotism and tradition on the website? |
| 31. | Do you think the government promotes harmony and consensus among their citizens on the website? |
| 32. | Do you think that massive activities and initiatives are promoted by the government? |
| 33. | Do you think the communication is directed towards the whole population? |
| 34. | Do you think the communication emphasis is directed towards tolerance and encouragement? |
| 35. | Do you think there is a lot of information about the staff and authorities displayed? |
| 36. | Do you think there is concern for the environment? |
| 37. | Do you think many women are in elected political positions? |
| 38. | Do you think there are certain activities that are more frequently directed towards men? |
| 39. | Do you think there seems to be a physical distance between the person of authority and others? |
| 40. | Do you think the government seems to be conservative, oriented towards law and order? |
| 41. | Do you think there is evidence of extremism and repression of extremism? |
| 42. | Do you think the information is presented in a simple and clear way? |
| 43. | Do you think the citizen has limited choices and the web site provides limited amount of data? |
| 44. | Do you think there is redundancy of information? |
| 45. | Do you think the information of the website is presented whit lots of graphics, charts and numbers? |

**Table 2 Proposed Items**

Finally, the proposed theoretical constructs derived from the literature review, and some of the proposed items, are presented in Table 3.

| | Power Distance | Uncertainty Avoidance | Masculinity/Femininity | Individualism/Collectivism |
|---|---|---|---|---|
| Layout | Website structure | Website | Consistency/flexibility | Density/Hierarchies |

| | | alignment | | |
|---|---|---|---|---|
| Navigation | List of contents | Consistency | Clarity of functions | Use of site maps |
| Design | Colors | Familiarity | Non-human images, human images | Graphics and tables |
| Content | Information organization and structure | Clarity and order of information<br>Availability | Audiences reached | Grouping of elements |

**Table 3 Proposed Theoretical Constructs**

## CONCLUSION

This document aims to present a research in progress effort. Ideally, most of the methods, instruments and specific research limitations should be identified and presented. For this particular task, we are open to receive all the feedback and complementary comment, suggestions and ideas that could help to the concretion of the task. Additionally, it is important to mention that the survey will suffer modifications, since the scale construction process is an iteration one. Finally, it is worth mentioning too that the process of trying to assess culture within the website usability evaluation is one that has captures the attention from several different disciplines, and HCI seems to be the discipline that has taken the matter into consideration in a more formal way by encouraging the future researchers to continue the previous works from several different authors that have approached the culturally oriented website usability evaluation phenomenon.

The research that we are undertaking here is designed as part of a larger program to examine the utility of universal site design, and to ascertain if highly localized sites (in both structure as well as content) are more likely to raise assessments of usability and aid in better decision-making on the part of web-site users.